\documentclass[twocolumn,showpacs,preprintnumbers,amsmath,amssymb,prl]{revtex4-1}

\usepackage{amsmath,amssymb}
\usepackage{graphicx}

\usepackage[dvips,colorlinks=true,bookmarks=false,citecolor=blue,urlcolor=blue]{hyperref} 

\newcommand{\ve}{\varepsilon}
\newcommand{\veh}{\varepsilon_h}
\newcommand{\vem}{\varepsilon_{\textnormal{Ag}}^{\textnormal{NL}}}
\newcommand{\om}{\omega}
\newcommand{\prp}{\scriptscriptstyle{\perp}}
\newcommand{\vrt}{\scriptscriptstyle{\Vert}}

\begin{document}

\title{Subwavelength modulational instability and plasmon oscillons in nanoparticle arrays}

\author{Roman E. Noskov$^{1}$,  Pavel A. Belov$^{1}$, and Yuri S. Kivshar$^{1,2}$}

\affiliation{$^{1}$National Research University of Information Technologies, Mechanics and Optics, St. Petersburg 197101, Russia \\
$^{2}$Nonlinear Physics Centre, Research School of Physics and Engineering, Australian National University, Canberra ACT 0200, Australia}

\begin{abstract}
We study modulational instability in nonlinear arrays of subwavelength metallic nanoparticles, and analyze numerically
nonlinear scenarios of the instability development.  We demonstrate that modulational instability can lead to the formation
of regular periodic or quasi-periodic modulations of the polarization. We reveal that such nonlinear nanoparticle arrays
can support long-lived standing and moving oscillating nonlinear localized modes -- {\em plasmon oscillons}.
\end{abstract}


\maketitle

Nonlinearity-induced instabilities are observed in many different branches of physics, and they provide probably
the most dramatic manifestation of strongly nonlinear effects that can occur in Nature. Modulational instability (MI)
in optics manifests itself in a decay of broad optical beams (or quasi–continuous wave pulses) into
optical filaments (or pulse trains)~\cite{MI_1,*MI_2,*MI_3,*MI_4}, and such effects are well documented in both theory
and experiment. MI is also observed for partially spatially incoherent light beams in noninstantaneous
nonlinear media with the pattern formation from noise~\cite{moti}. It is expected that the study of subwavelength nonlinear
systems such as metallic nanowires or nanoparticle arrays may bring many new features
to the physics of MI and the scenarios of its development, however such effects were never studied before.

Over the past decade, surface plasmon polaritons (or plasmons) were suggested as the mean to overcome the diffraction limit in optical systems. In particular, by using plasmons excited in a chain of resonantly coupled metallic nanoparticles~\cite{ol,*prl}, one can spatially confine and manipulate optical energy over distances much smaller than the wavelength. In addition, strong geometric confinement can boost efficiency of nonlinear optical effects, including the existence of subwavelength solitons~\cite{prl_zhang,*prl_panoiu}.

In this Letter, we study modulational instability in subwavelength nonlinear systems for an array of optically driven metallic
nanoparticles~\cite{Yong,prb_0,citrin,prb_1} with a nonlinear response. We demonstrate the existence of novel types
of nonlinear effects in such subwavelength systems never discussed before, including the generation of regular or quasi-periodic polarization patterns and oscillating localized modes which can be termed {\em oscillons}, in analogy with the similar localized modes excited in driven granular materials~\cite{oscillon} and Newtonian fluids~\cite{Arbell}.

\begin{figure}[b]
\centerline{\mbox{\resizebox{8.5cm}{!}{\includegraphics{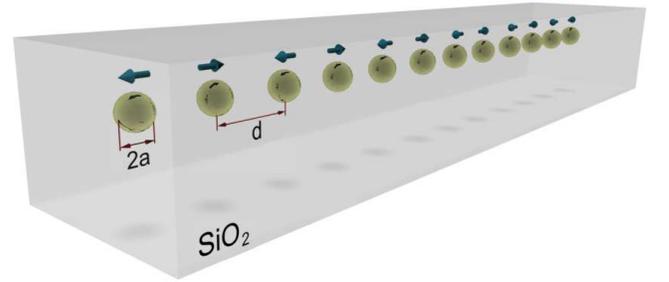}}}}
\caption{\label{fig:1} (Color online) Schematic sketch demonstrating geometry of the studied problem.
Arrows indicate particle polarizations after MI development.}\label{fig1}
\end{figure}

Figure~\ref{fig1} shows the geometry of our problem: a chain of identical spherical silver nanoparticles is embedded into a fused silica host medium with permittivity $\veh$ and driven by an external optical field with the frequency close to the frequency of the surface plasmon resonance of an individual particle. We assume that the particle radius and distance between the particles are $a=10$~nm and $d=30$~nm, respectively. Ratio $a/d$ satisfies the condition $a/d \leq1/3$, so that we can employ
the point dipole approximation~\cite{Yong}. In the optical spectral range, a linear part of silver dielectric constant can be written in a generalized Drude form $\ve_{\textnormal{Ag}}^\textnormal{L}=\ve_\infty-\om_p^2/[\om(\om-i\nu)]$, where $\ve_\infty=4.96$, $\hbar\om_p=9.54$~eV, $\hbar\nu=0.55$~eV~\cite{Johnson} (hereinafter we accept $\exp(i\om t)$ time dependence); whereas dispersion of SiO$_2$ is neglected since $\veh\simeq2.15$ for wavelengths 350–-450~nm~\cite{Palik}. Nonlinear dielectric constant of silver is $\ve_{\textnormal{Ag}}^{\textnormal{NL}}=\ve_{\textnormal{Ag}}^\textnormal{L}+\chi^{(3)}|{\bf E}^{(in)}_n|^2$, where ${\bf E}^{(in)}_n$ is the local field inside $n$-th particle. We keep only cubic susceptibility due to spherical symmetry of particles. Currently, there is no reliable theoretical models describing nonlinear optical response of metal nanoparticles,
however experimental data shows that $\chi^{(3)}$ depends on many factors, including duration and frequency of the external excitation as well as particle characteristics themselves (metal type and size)~\cite{Palpant}. According to the model suggested in Ref.~\cite{Drachev} and confirmed in experiment, 10 nm radii Ag spheres possess a remarkably high and purely real cubic susceptibility $\chi^{(3)}\simeq 3\times 10^{-9}$~esu, in comparing to which the cubic nonlinearity of SiO$_2$ is weak ($\thicksim10^{-15}$~esu~\cite{weber_book_03}).

We study nonlinear dynamics of our chain by employing the dispersion relation method~\cite{Whitham} that allows deriving a system of coupled equations for slowly varying amplitudes of the particle dipole moments. This approach is based on the assumption that in the system there are small and large time scales, which in our case is fulfilled automatically since each particle acts as a resonantly excited oscillator with slow (in comparison with the light period) inertial response.

We start with the standard expression for the electric dipole moment
induced in the $n$-th particle written for Fourier transforms
\begin{equation}
\label{n-th dipole moment} \alpha_n(\om)^{-1}{\bf p}_{n}={\bf E}_{n}^{(ex)}+\sum_{m \neq n}{\bf E}_{n,m},
\end{equation}
where $$\alpha_n(\om)=\veh\left\{\frac{\vem(\om)+2\veh}{a^3[\vem(\om)-\veh]}+i \frac{2}{3} k^3\right\}^{-1}$$ is the electric polarizability of
the $n$-th particle, ${\bf E}_{n}^{(ex)}$ is the external electric field acting on $n$-th particle,
\begin{multline*}
{\bf E}_{n,m}= \left( \left( 1+ikd|n-m| \right) \frac{3({\bf r}_0\cdot{\bf p}_m){\bf r}_0-{\bf p}_m}{\veh|n-m|^3d^3} \right. {} \\
{} \left. +k^2 \frac{{\bf p}_m - ({\bf r}_0 \cdot{\bf p}_m){\bf r}_0}{\veh|n-m| d} \right)e^{-ikd|n-m|}
\end{multline*}
is in charge of dipole-dipole interaction between $m$-th and $n$-th particles, $k=\om/c\sqrt{\veh}$, ${\bf r}_0$ is the unit vector pointing from
the $m$-th to the $n$-th particle. Assuming that $\chi^{(3)} |{\bf E}^{(in)}_n|^2 \ll 1$ and $\nu/\om_0 \ll 1$, we decompose $\alpha_n(\om)^{-1}$
in the vicinity of the frequency of the surface plasmon resonance of an individual particle, $\om_0=\om_p/\sqrt{\ve_\infty + 2\veh}$, and keep the first-order terms involving time derivatives for describing (actually small) broadening of the particle polarization spectrum,
\begin{equation}
\label{polarizability} \alpha_n^{-1} \approx \alpha_n(\om_0)^{-1} + \left. \frac{d\alpha_n^{-1}}{d\om} \right|_{\om=\om_0} \left(\Delta \om-i\frac{d}{dt}\right),
\end{equation}
where $\Delta \om$ is the frequency shift from the resonance value. Having expressed ${\bf E}^{(in)}_n$ via ${\bf p}_n$, we
substitute Eq. (\ref{polarizability}) into Eq. (\ref{n-th dipole moment}) and write ${\bf E}_{n,m}$ in the same order of the perturbation theory and obtain the equations,
\begin{equation}\label{dynamic}
\begin{split}
-i\frac{d P_n^{\prp}}{d\tau}+\left(-i\gamma+\Omega+|{\bf P}_n|^2 \right) P_n^{\prp}+ \sum_{m\neq n} G_{n,m}^{\prp} P_m^{\prp} &= E_n^{\prp}, \\
-i\frac{d P_n^{\vrt}}{d\tau}+\left(-i\gamma+\Omega+|{\bf P}_n|^2 \right) P_n^{\vrt}+ \sum_{m\neq n} G_{n,m}^{\vrt} P_m^{\vrt} &= E_n^{\vrt},
\end{split}
\end{equation}
where
$$
G_{n,m}^{\prp}  = \frac{\eta}{2} \left( (k_0 d)^2 - \frac{i k_0 d}{|n-m|}- \frac{1}{|n-m|^2} \right) \frac{e^{-i k_0 d|n-m|}}{|n-m|}, $$
$$
G_{n,m}^{\vrt}  = \eta \left(\frac{i k_0 d}{|n-m|} + \frac{1}{|n-m|^2} \right) \frac{e^{-i k_0 d|n-m|}}{|n-m|},
$$
$P_n^{\prp,\vrt}=p_n^{\prp,\vrt}\sqrt{\chi^{(3)}}/(\sqrt{2(\ve_\infty+2\veh)}\veh a^3)$ and $E_n^{\prp,\vrt} = -3 \veh \sqrt{\chi^{(3)}} E^{(ex)\prp,\vrt}_n/\sqrt{8(\ve_\infty+2\veh)^3}$ are dimensionless slowly varying amplitudes of the particle dipole moments and external electric field,
 respectively, the indices '$\prp$' and '$\vrt$' stand for the transverse and longitudinal components with respect to the chain axis,
$\eta=\frac{3\veh}{\ve_{\infty}+2\veh}\left(\frac{a}{d}\right)^3$, $|{\bf P}_n|^2=|P_n^{\prp}|^2+|P_n^{\vrt}|^2$,  $\gamma=\nu/(2\om_0)+(k_0 a)^3\veh/(\ve_\infty+2\veh)$ describes thermal and radiation losses of particles, $k_0=\om_0/c\sqrt{\veh}$, $\Omega=(\om-\om_0)/\om_0$ and $\tau = \om_0 t$. Equations~(\ref{dynamic}) describe
temporal nonlinear dynamics of a chain of metallic nanoparticles driven by arbitrary external optical field with the frequency $\om\sim\om_0$. We stress that the suggested model takes into account all particle interactions through the dipole fields, and it can be applied to both finite and infinite chains, being also extended to higher dimensions.

\begin{figure}
\centerline{\mbox{\resizebox{8.3cm}{!}{\includegraphics{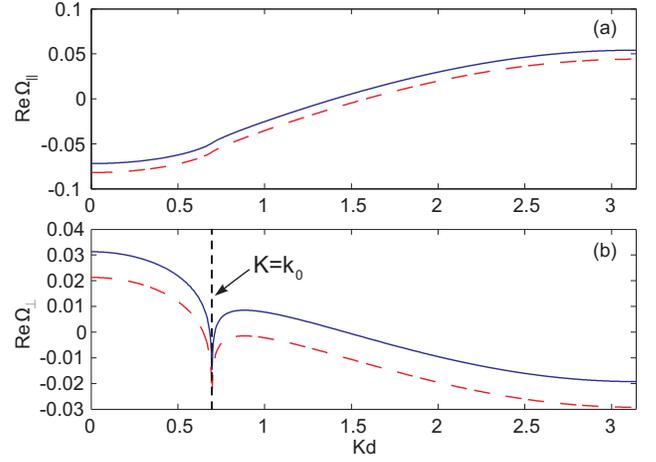}}}}
\caption{(Color online) \label{fig:2} Nonlinear dispersions of (a) longitudinal and (b) transverse eigenmodes
of an infinite chain. Dashed curves correspond to the linear limit. Vertical dashed line in (b) marks the light line,
$k_0=\om_0/c\sqrt{\veh}$.}
\label{fig2}
\end{figure}

First, we consider an infinite chain. For the stationary unbiased linear case, when $d/d\tau =0$ and $E_n^{\prp,\vrt}=0$, we look for solutions in the form $P_n^{\prp,\vrt}\sim\exp{(-inKd)}$, and from Eqs. (\ref{dynamic}) find well-known dispersion relations for transverse and longitudinal eigenmodes of the system~\cite{citrin}, shown in Fig.~\ref{fig2}. Taking into account nonlinearity just shifts the dispersion curves along the frequency axis. Light line, which for $\Omega<<1$ takes the form $k_0=\om_0/c\sqrt{\veh}$, divides the eigenmodes into fast (with $K<k_0$) and slow (with $K>k_0$) experienced strong and weak radiation damping, respectively. Logarithmic singularity occurred at $K=k_0$ for the transverse modes is caused by the phase matching between the chain mode and the plane wave traveling in the host medium.

To study MI, we excite the chain by an homogenous electric field with one of the two polarizations: (i) ${\bf E}_n=(E_0^{\prp},0)$ and (ii) ${\bf E}_n=(0,E_0^{\vrt})$. In this case, all particle dipole moments remain the same, $P_n^{\prp,\vrt}=P_0^{\prp,\vrt}$, and the system stationary states can be written as follows
\begin{equation}\label{initial state}
\left(-i\gamma+\Omega+\sum_{j=1}^\infty A_j^{\prp,\vrt}+|P_0^{\prp,\vrt}|^2 \right) P_0^{\prp,\vrt}= E_0^{\prp,\vrt},
\end{equation}
where $A_j^{\prp}=\eta\left((k_0 d)^2 j^{-1}-i k_0 d j^{-2} -j^{-3} \right) \exp{(-ik_0 d j)},$ and $A_j^{\vrt}=2\eta\left(i k_0 d j^{-2}+j^{-3} \right) \exp{(-ik_0 d j)}$. Transition from $G_{n,m}^{\prp,\vrt}$ to $A_j^{\prp,\vrt}$ has been made via the replacement $|n-m|=j$ and taking into account symmetry structure of the series. When $\Omega<-{\rm Re}\sum_{j=1}^\infty A_j^{\prp,\vrt} - \sqrt{3}\left( \gamma - {\rm Im}\sum_{j=1}^\infty A_j^{\prp,\vrt} \right)$, the polarization $P_0^{\prp,\vrt}$ becomes a three-valued function of $E_0^{\prp,\vrt}$ leading to bistability.

Next, we analyze linear stability of the stationary states with respect to weak spatiotemporal modulations and derive the expression for the instability growth rate,
\[
\label{increment}
\lambda_{\prp,\vrt}=\tilde{\gamma}_{\prp,\vrt} + \biggr\{|P_0^{\prp,\vrt}|^4-\biggr(2|P_0^{\prp,\vrt}|^2+\Omega+{\rm{Re}} \sum_{j=1}^\infty B_j^{\prp,\vrt}\biggr)^2\biggr\}^{1/2},
\]
where $\tilde{\gamma}_{\prp,\vrt}={\rm{Im}} \sum_{j=1}^\infty B_j^{\prp,\vrt} - \gamma$, $B_j^{\prp,\vrt}=A_j^{\prp,\vrt} \cos(Kdj)$. Thus, the initial nonlinear homogenous states (\ref{initial state}) become unstable provided $\lambda_{\prp,\vrt}>0$. The stability depends on the external field parameters $E^{\prp,\vrt}_0$ and $\Omega$ as well as on the modulation wavenumber $K$.

\begin{figure}
\centerline{
\mbox{\resizebox{8.5cm}{!}{\includegraphics{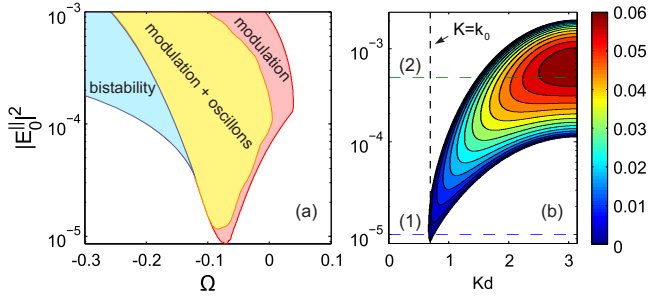}}}}
\caption{(Color online)\label{fig:3} (a) Bifurcation diagram showing bistability regime and different scenarios of modulation instability development for longitudinal excitations, as a function of $\Omega$ and $|E^{\vrt}_0|^2$. (b) Contour map of $\lambda_{\vrt}$ on the plane $(Kd,|E^{\vrt}_0|^2)$ at $\Omega=-0.07$. Horizontal dashed lines (1) and (2) mark the intensities of the external light used in numerical simulations of Eq.~(\ref{dynamic}) shown in Figs.~\ref{fig4}(a,b), respectively.}
\label{fig3}
\end{figure}

Next, we consider the case of the longitudinal excitation in detail. The condition $\lambda_{\vrt}=0$ at any $K$ defines the boundaries of MI in the plane ($\Omega, |E^{\vrt}_0|^2$) shown in Fig. \ref{fig3}(a). Interestingly, the middle and upper branches in the bistable region of dependency $P_0^{\vrt}(E_0^{\vrt})$ also correspond to MI, but the development inside the bistability region cannot be reached because the middle branch is unstable, while the system transition from the lower to upper branch itself initiates appearance of MI.

Figure~\ref{fig3}(b) shows a contour map of $\lambda_{\vrt}$ in the plane $(Kd,|E^{\vrt}_0|^2)$ at $\Omega=-0.07$. Remarkably, MI takes place only for slow eigenmodes of the chain. As follows from Fig. \ref{fig3}(b), one can manage eigenmode spectrum excited during MI growth by varying $|E^{\vrt}_0|^2$ only. In particular, when $|E^{\vrt}_0|^2$ is chosen to be close to the lower or upper edge of the MI domain, just one spatial harmonic should be excited,  with correspondingly $Kd\simeq0.7$ or $Kd=\pi$.

\begin{figure}
\centerline{
\mbox{\resizebox{8.5cm}{!}{\includegraphics{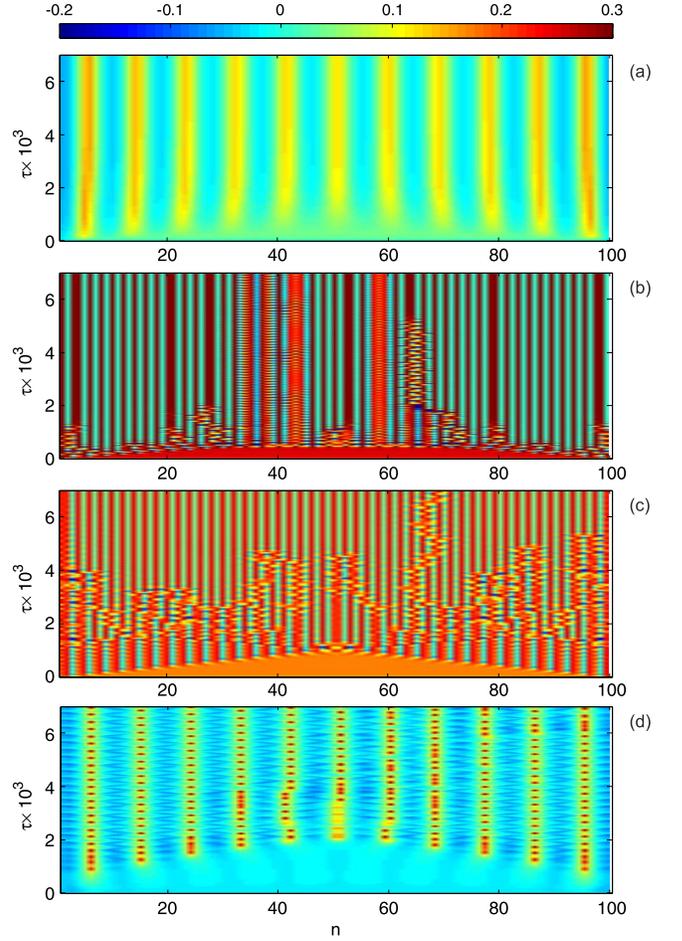}}}}
\caption{(Color online)\label{fig:4} Dynamics of Re~$P_n^{\vrt}$ obtained by numerical simulations of Eq.~(\ref{dynamic}) for a finite chain excited longitudinally with (a) $\Omega=-0.07$, $|E^{\vrt}_0|^2=0.1\times10^{-4}$; (b) $\Omega=-0.07$, $|E^{\vrt}_0|^2=4.9\times10^{-4}$; (c) $\Omega=-0.02$, $|E^{\vrt}_0|^2=2.65\times10^{-4}$; and (d) $\Omega=-0.09$ and $|E^{\vrt}_0|^2=0.11\times10^{-4}$.}
\label{fig4}
\end{figure}

However, the linear stability analysis does not provide any information about the subsequent evolution of the unstable system, especially
when the external field excites a broad spectrum of eigenmodes. To analyze those scenarios, we perform numerical simulations of Eq. (\ref{dynamic}) for a finite chain (with 100 nanoparticles) at zero initial conditions. Edge effects play a role of small perturbations needed for generating MI. The amplitude of the homogeneous external field is supposed to be slowly growing to the saturation level $E^{\vrt}_0$ (which is reached at $\tau\approx100$) lying in the MI zone.

Characteristic results are summarized in Figs.~\ref{fig4}(a-d). When $E^{\vrt}_0$ crosses the lower edge of the MI domain [depicted by a dashed line (1) in Fig.~\ref{fig3}(b)], we observe that MI results in the excitation of one eigenmode with $Kd\simeq0.7$ [see Fig.~\ref{fig4}(a)], in accord with the prediction of the linear stability analysis. The excited eigenmode acts as modulation of the initial almost homogenous state which becomes unstable. That is why Re~$P_n^{\vrt}$ tend to be predominantly positive. They are just biased by the external field.
\begin{figure}
\centerline{
\mbox{\resizebox{8.3cm}{!}{\includegraphics{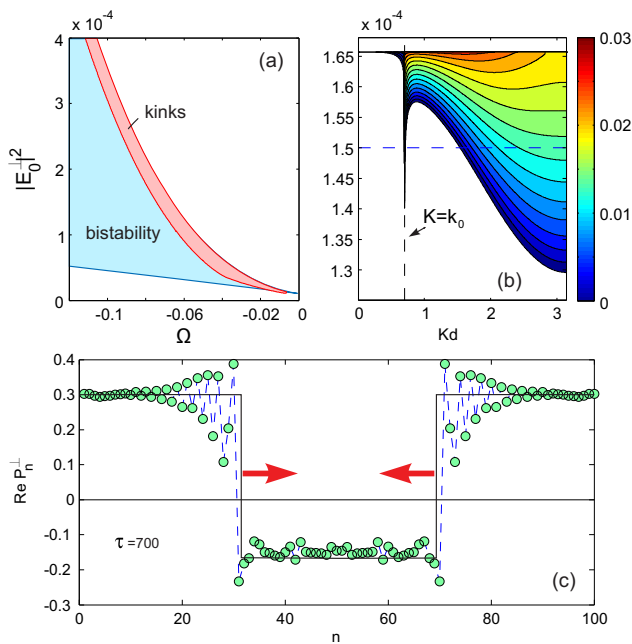}}}}
\caption{(Color online)\label{fig:5} (a) Bifurcation diagram for transverse excitation. (b) Contour map of $\lambda_\perp$ on the plane of parameters $(Kd,|E^{\perp}_0|^2)$ at $\Omega=-0.07$. (c) Snapshot of Re~$P_n^{\perp}$ at $\tau=700$ obtained numerically from Eq. (\ref{dynamic}) with $\Omega=-0.07$ and $|E^{\prp}_0|^2=1.5\times10^{-4}$, indicated in (b) by a horizontal dashed line. Points joined by dashed lines give a guide for the eye.}
\label{fig5}
\end{figure}

Figure~\ref{fig4}(b) shows the case when the external field of larger amplitude excites a wide eigenmode spectrum [indicated by dashed line (2) in Fig. \ref{fig3}(b)] \cite{SM}. Here, MI leads to the formation of a stationary higher-order mode along with oscillating localized states. Some of them appeared to be unstable and decay, whereas others remain stable. Importantly, such soliton-like localized modes may be at rest or they can drift slowly along the chain, as shown in Figs.~\ref{fig4} (b,c). We notice that these oscillatory localized states in a driven chain are very similar to spatiotemporal structures termed {\em oscillons}  observed previously in other types of dissipative systems~\cite{oscillon,Arbell}, and we refer to them as {\it plasmon oscillons}. We point out that the plasmon oscillons may exist not only in the form of solitary states but they also can create patterns, as illustrated in Fig.~\ref{fig4}(d) \cite{SM}. We have studied some of the properties of such oscillatory states, and the results will be published elsewhere.

Finally, we conduct the similar analysis for the case of the transversal excitation. Figure~\ref{fig5}(a) shows the corresponding bifurcation diagram. In contrast to the longitudinal case, the MI region is fully placed inside the bistability domain capturing a part of the lower branch in the dependence $P_0^{\prp}(E_0^{\prp})$. According to the contour map of $\lambda_\perp$ in the plane $(Kd,|E^{\prp}_0|^2)$ shown in Fig.~\ref{fig5}(b), the spectrum of excited eigenmodes can be tuned by varying the value of $|E^{\prp}_0|^2$, in analogy with the longitudinal case. Nevertheless, the width of spectrum weakly affects the scenarios of the MI development. Numerical simulations of Eq. (\ref{dynamic}) demonstrate that, independently on the value of $E^{\prp}_0$, the growth of MI results in switching of the system from the lower to upper branch in the bistability region of $P_0^{\prp}(E_0^{\prp})$. In the case of a finite chain, MI is accompanied by a pair of switching waves (kinks) at the edges which move towards each other, as shown in  Fig.~\ref{fig5}(c) \cite{SM}.

The observation of MI requires high illuminating powers, higher than 10 MW/cm$^2$, that could cause thermal damage to particles. To estimate maximal duration of the external laser pulse, we use the results of previous studies on the ablation thresholds for gold films~\cite{pronko_jap_1995,*gamaly} providing values of 1.6 J/cm$^2$ and 0.6 J/cm$^2$ for 1 ns and 1 ps pulses, respectively. Gold demonstrates stronger thermal losses than silver at optical frequencies. That is why this data is completely acceptable. Taking into account amplification of electric field inside nanoparticles due to surface plasmon resonance, we come to the external threshold intensities of 3.6 MW/cm$^2$ and 1.3 GW/cm$^2$ corresponding to 1 ns and 1 ps pulses, respectively. Thus, ablation of silver particles will not be critical at least till pulse durations of 1 ps. As the characteristic time of the MI growth is of $(\lambda_{\prp,\vrt}\om_0)^{-1}\simeq 10$~fs that is much less than the maximal pulse duration, all predicted effects seem readily observable in experiment.

In conclusion, we have studied theoretically modulational instability in arrays of subwavelength metallic nanoparticles,
and analyzed numerically the development of such instabilities beyond the linear approximation.
We have observed that modulational instability can be enhanced substantially by the geometric confinement, and it can lead to the formation of regular periodic or quasi-periodic polarization patterns. We have observed the generation
of long-lived standing and moving oscillating nonlinear localized modes in the form  of plasmon oscillons.
The experimental observation of the predicted modulational instability can provide a prominent approach to achieve
subwavelength confinement of the optical fields guided by plasmonic nanostructures.

The authors acknowledge a support from the Australian Research Council and a mega-grant of the Ministry of Education and Science of Russian Federation, as well as fruitful discussions with A.A.~Zharov.

\bibliography{References}
\end{document}